\documentclass{JHEP3}
\usepackage{amsmath}
\usepackage{amssymb}
\usepackage{graphicx}
\usepackage{cite}
\title{Abelian monopole and vortex condensation in lattice gauge theories}

\author{Paolo Cea \\ Dipartimento
Interateneo di Fisica, Universit\`a di Bari  and INFN - Sezione di Bari, \\
I-70126 Bari, Italy \\
E-mail:  \email{Paolo.Cea@ba.infn.it} }

\author{Leonardo Cosmai \\ INFN - Sezione di Bari, I-70126 Bari,
Italy\\
E-mail:  \email{Leonardo.Cosmai@ba.infn.it}}

\received{\today}

\preprint{BARI-TH 408/2001}

\abstract{We study Abelian monopole and vortex condensation in
lattice pure gauge theories. Condensation is detected by means of
a disorder parameter defined in terms of a gauge-invariant
effective action introduced using the lattice Schr\"odinger
functional. Dirac monopoles condense in the confined phase of U(1)
lattice gauge theory. Abelian monopoles and Abelian vortices
condense in the confined phase of SU(2) and SU(3) lattice gauge
theories.}

\keywords{Confinement, Lattice Gauge Field Theories}

\begin{document}

\section{Introduction}

To give a possible explanation of color confinement
G.~'t~Hooft~\cite{tHooft:1976eps} and
S.~Mandelstam~\cite{Mandelstam:1974pi} suggested long time ago
that vacuum in gauge theories behaves like a magnetic (dual)
superconductor. The dual superconductivity hypothesis relies upon
the very general assumption that the dual superconductivity of the
ground state is realized if there is condensation of Abelian
magnetic monopoles. Indeed, lattice calculations have given
evidence of Abelian magnetic monopoles
condensation~\cite{DiGiacomo:1994jy,DiGiacomo:1999fa,DiGiacomo:1999fb,Carmona:2001ja,
Cea:2000zr,Shiba:1995db,Arasaki:1997sm,Nakamura:1997sw,Chernodub:1997ps,Jersak:1999nv,Hoelbling:2000su}.

On the other hand numerical evidence has
emerged~\cite{Faber:1999gu,Langfeld:1998cz,deForcrand:1999ms,
Bertle:1999vu,Ambjorn:1999ym,Montero:1999gq,Jahn:1999ab,
Stack:2000zx,Bakker:1999qh,Alexandru:2000ir,DelDebbio:2000cx,deForcrand:2000pg,Kovacs:2000sy,Korthals-Altes:2000gs}
in favor of the so called center vortex picture where the vacuum
consists of a coherent condensate of magnetic flux tubes. Also
this theoretical proposal has been advanced long time
ago~\cite{'tHooft:1978hy,Cornwall:1979hz,Cornwall:1998ds,Yaffe:1980iq,Mack:1979rq,Mack:1980kr,Tomboulis:1981vt,Tomboulis:1993wm}.

In this paper we compare the dual superconductivity scenario with
the vortex condensate picture both in Abelian and non Abelian pure
gauge lattice theories. A partial account of results discussed in
the present paper has been reported in Ref.~\cite{Cea:2000rj}. In
this work we do not consider the center vortices. Instead we study
the Abelian vortices which eventually give rise to a coherent
condensate of Abelian magnetic flux tubes.

Monopole or vortex condensation can be detected using
order/disorder
parameters~\cite{Kadanoff:1971kz,Fradkin:1978th,Davis:2000kv}. At
zero temperature we employ a disorder parameter defined in terms
of a gauge-invariant lattice effective action
introduced~\cite{Cea:1997ff,Cea:1997ku,Cea:1999gn} using the
lattice Schr\"odinger
functional~\cite{Luscher:1992an,Luscher:1995vs,Sint:1994un,Luscher:1993zx}.
At finite temperature the disorder parameter has been
defined~\cite{Cea:2000zr} in terms of a thermal partition
functional.

For the sake of clarity let us recall our implementation of the gauge-invariant lattice
Schr\"odinger functional previously discussed in Ref.~\cite{Cea:1999gn}.
Let us consider the continuum Euclidean Schr\"odinger functional in Yang-Mills
theories without matter field:
\begin{equation}
\label{Zeta} {\mathcal{Z}}[\mathbf{A}^{(f)},\mathbf{A}^{(i)}] =
\left\langle \mathbf{A}^{(f)} \left| e^{-HT} {\mathcal{P}} \right|
\mathbf{A}^{(i)} \right\rangle \,.
\end{equation}
In Eq.~(\ref{Zeta}) $H$ is the pure gauge Yang-Mills Hamiltonian
in the temporal gauge, $T$ is the Euclidean time extension, while
${\mathcal{P}}$ projects onto the physical states.
$A^{a(i)}_k(\vec{x})$ and $A^{a(f)}_k(\vec{x})$ are static
classical gauge fields, and the state $|\mathbf{A}\rangle$ is such
that
\begin{equation}
\label{stateA} \langle \mathbf{A} | \Psi \rangle =
\Psi[\mathbf{A}] \,.
\end{equation}
Inserting an orthonormal basis $|\Psi_n\rangle$ of gauge invariant
energy eigenstates in Eq.~(\ref{Zeta})
\begin{equation}
\label{Zetaortho} {\mathcal{Z}}[\mathbf{A}^{(f)},\mathbf{A}^{(i)}]
= \sum_n e^{-E_n T} \Psi_n[\mathbf{A}^{(f)}]
\Psi^{*}[\mathbf{A}^{(i)}] \,.
\end{equation}
Since we are interested in the lattice version of the
Schr\"odinger functional, it makes sense to perform a discrete sum
in Eq.~(\ref{Zetaortho}) for the spectrum is discrete in a finite
volume. Eq.~(\ref{Zetaortho}) shows that the Schr\"odinger
functional is invariant under arbitrary static gauge
transformations of the fields $\mathbf{A}^{(f)}$ and
$\mathbf{A}^{(i)}$.

Using standard formal manipulations and the gauge invariance of
the Schr\"odinger functional it is easy to rewrite
${\mathcal{Z}}[\mathbf{A}^{(f)},\mathbf{A}^{(i)}]$ as a functional
integral~\cite{Rossi:1980jf,Rossi:1980pg,Gross:1981br}
\begin{equation}
\label{Zetaint} {\mathcal{Z}}[\mathbf{A}^{(f)},\mathbf{A}^{(i)}] =
\int {\mathcal{D}}A_\mu \; e^{-\int_0^T dx_4 \, \int d^3x \,
{\mathcal{L}}_{YM}(x)}  \,,
\end{equation}
with the constraints:
\begin{eqnarray}
\label{constraints}
\mathbf{A}(x_0=0) = & \mathbf{A}^{(i)} \,, \nonumber \\
\\
\mathbf{A}(x_0=T) = & \mathbf{A}^{(f)} \,. \nonumber
\end{eqnarray}
Strictly speaking we should include in Eq.~(\ref{Zetaint}) the sum
over topological inequivalent classes. However, it turns out
that~\cite{Luscher:1992an,Luscher:1995vs,Sint:1994un,Luscher:1993zx} on the lattice such an
average is not needed because the functional integral
Eq.~(\ref{Zetaint}) is already invariant under arbitrary gauge
transformations of $\mathbf{A}^{(i)}$ and $\mathbf{A}^{(f)}$.

On the lattice the natural relation between the continuum gauge fields and
the corresponding lattice links is given by
\begin{equation}
\label{links}
U_\mu(x) = {\mathrm P} \exp\left\{
iag \int_0^1 dt \, A_\mu(x+at\hat{\mu}) \right\}
\end{equation}
where ${\mathrm P}$ is the path-ordering operator, $a$ is the
lattice spacing and $g$ the bare gauge coupling constant.

The lattice
implementation~\cite{Luscher:1992an,Luscher:1995vs,Sint:1994un,Luscher:1993zx}
of the Schr\"odinger functional, Eq.~(\ref{Zetaint}), is now
straightforward:
\begin{equation}
\label{Zetalatt} {\mathcal{Z}}[U^{(f)},U^{(i)}] = \int
{\mathcal{D}}U_\mu \; e^{-S} \,.
\end{equation}
In Eq.~(\ref{Zetalatt}) the functional integration is done over
the links $U_\mu(x)$ with the fixed boundary values:
\begin{equation}
\label{boundary} U_k(x)|_{x_4=0} = U_k^{(i)}\,, \quad
U_k(x)|_{x_4=T} = U_k^{(f)} \,\,\,\,;\,\,\, (k=1,2,3)\,.
\end{equation}
Links in temporal direction are not constrained.
$S$ is the standard Wilson action
modified~\cite{Luscher:1992an,Luscher:1995vs,Sint:1994un,Luscher:1993zx} to take into account
the boundaries at $x_4=0,T$. For SU(N)
\begin{equation}
S = \beta \sum_{x,\mu>\nu} W_{\mu\nu}(x) \, \left\{ 1 -
\frac{1}{N} \mathrm{Re} \, \mathrm{Tr}[1 - U_{\mu\nu}(x)] \right\}
\,, \qquad \beta = \frac{2N}{g^2}  \,,
\end{equation}
where $U_{\mu\nu}(x)$ are the plaquettes in the $(\mu,\nu)$-plane
and the weights $W_{\mu\nu}(x)$ are given by
\begin{equation}
\label{Wmunu}
W_{\mu\nu}(x) = \left\{
\begin{array}{ll}
1/2  &  \mbox{spatial plaquettes at $x_4=0,T$} \\
1    &  \mbox{otherwise}
\end{array}
\right. \,.
\end{equation}
It is possible to improve the lattice action $S$ by modifying
the weights $W_{\mu\nu}$'s~\cite{Luscher:1992an,Luscher:1995vs,Sint:1994un,Luscher:1993zx}.
Note that, due to the fact that $U^{(i)} \ne U^{(f)}$, one cannot impose
periodic boundary conditions in the Euclidean time direction. On the other hand
one can assume periodic boundary conditions in the spatial directions.

Let us consider, now, a static external background field
$\mathbf{A}^{\mathrm{ext}}(\vec{x}) =
\mathbf{A}^{\mathrm{ext}}_a(\vec{x}) \lambda_a/2$, where
$\lambda_a/2$ are the generators of the SU(N) Lie algebra. We
introduce a new functional:
\begin{equation}
\label{Gamma} \Gamma[\mathbf{A}^{\mathrm{ext}}] = -\frac{1}{T} \ln
\left\{ \frac{{\mathcal{Z}}[U^{\mathrm{ext}}]}{{\mathcal{Z}}[0]}
\right\} \,,
\end{equation}
where
\begin{equation}
\label{ZetaUext}
{\mathcal{Z}}[U^{\mathrm{ext}}] = {\mathcal{Z}}[U^{\mathrm{ext}},U^{\mathrm{ext}}] \,,
\end{equation}
and ${\mathcal{Z}}[0]$ is the Schr\"odinger functional
Eq.~(\ref{ZetaUext}) with
$\mathbf{A}^{\mathrm{ext}}=\boldsymbol{0}$
($U^{\mathrm{ext}}_k={\mathbf 1}$). The lattice link
$U_k^{\mathrm{ext}}$ is obtained from the continuum background
field $\mathbf{A}^{\mathrm{ext}}$ through Eq.~(\ref{links}).

From the previous discussion it is clear that
$\Gamma[\mathbf{A}^{\mathrm{ext}}]$ is invariant for lattice gauge
transformations of the external links $U^{\mathrm{ext}}_k$.
Moreover, from Eq.~(\ref{Zetaortho}) it follows that
\begin{equation}
\label{limit} \lim_{T \to \infty}
\Gamma[\mathbf{A}^{\mathrm{ext}}] = E_0[\mathbf{A}^{\mathrm{ext}}]
- E_0[0]
\end{equation}
where $E_0[\mathbf{A}^{\mathrm{ext}}]$ is the vacuum energy in
presence of the external background field. In other words
$\Gamma[\mathbf{A}^{\mathrm{ext}}]$ is the lattice gauge-invariant
effective action for the static background field
$\mathbf{A}^{\mathrm{ext}}$.

Note that, since our definition of the lattice effective action
uses the lattice Schr\"odinger functional with the same boundary
fields at $x_4=0$ and $x_4=T$, we can glue the two hyperplanes
$x_4=0$ and $x_4=T$ together. This way we end up in a lattice with
periodic boundary conditions in time direction too. Therefore our
lattice Schr\"odinger functional turns out to be
\begin{equation}
\label{Zetal} {\mathcal{Z}}[U_k^{\mathrm{ext}}] = \int
{\mathcal{D}}U_\mu \; e^{-S} \;,
\end{equation}
where the functional integral is defined over a four-dimensional
hypertorus with the "cold-wall"
\begin{equation}
\label{coldwall} U_k(x_4,\vec{x})|_{x_4=0} =
U^{\mathrm{ext}}_k(\vec{x})  \,.
\end{equation}
Moreover, due to the lacking of free boundaries, the lattice
action in Eq.~(\ref{Zetal}) is now the familiar Wilson action
\begin{equation}
\label{SWilson} S_W = \beta \sum_{x,\mu>\nu} \, \left\{ 1 -
\frac{1}{N} \mathrm{Re} \, \mathrm{Tr}[1 - U_{\mu\nu}(x)] \right\}
\,.
\end{equation}
We also impose that links at the spatial boundaries (we mean the
spatial boundaries of each given time slice and not only spatial
boundaries of the slice $x_4=0$) are fixed according to
Eq.~(\ref{coldwall}). In the continuum this last condition amounts
to the usual requirement that the fluctuations
over the background field vanish at  infinity. \\

We want to extend our definition of lattice effective action to
gauge systems at finite temperature. In this case the relevant
quantity is the thermal partition function. In the continuum we
have:
\begin{equation}
\label{thermal} \text{Tr}\left[ e^{-\beta_T H} \right] = \int
\mathcal{D}\mathbf{A} \, \langle \mathbf{A} \left | e^{-\beta_T H}
\mathcal{P} \right| \mathbf{A} \rangle \,,
\end{equation}
where $\beta_T$ is the inverse of the physical temperature,
$H$ is the Hamiltonian, and $\mathcal{P}$ projects onto the
physical states. As is well known, the thermal partition function
can be written as~\cite{Gross:1981br}:
\begin{equation}
\label{tpf}
\text{Tr}\left[ e^{-\beta_T H} \right] =
\int_{A_\mu(\beta_T,\vec{x})=A_\mu(0,\vec{x})}
 \mathcal{D}A_\mu(x_4,\vec{x})  \,
e^{-\int^{\beta_T}_0 dx_4 \, \int d^3 \vec{x} \mathcal{L}_{Y-M}(\vec{x},x_4)}
\,.
\end{equation}
On the lattice we have:
\begin{equation}
\label{tpflattice} \text{Tr}\left[ e^{-\beta_T H} \right] =
\int_{U_\mu(\beta_T,\vec{x})=U_\mu(0,\vec{x})}
 \mathcal{D}U_\mu(x_4,\vec{x})  \,
e^{-S_W} \,.
\end{equation}
Comparing Eq.~(\ref{tpflattice}) with Eqs.~(\ref{Zetal})
and~(\ref{coldwall}), we get:
\begin{equation}
\label{trace} \text{Tr}\left[ e^{-\beta_T H} \right] = \int
\mathcal{D}U_k(\vec{x})  \, \mathcal{Z}[U_k(\vec{x})] \,,
\end{equation}
where $\mathcal{Z}[U_k(\vec{x})]$ is the Schr\"odinger functional
Eq.~(\ref{Zetal}) defined on a lattice with $L_4=\beta_T$, with
"external" links $U_k(\vec{x})$ at $x_4=0$. We recall that, by
definition, $\mathcal{Z}[U_k(\vec{x})]$ includes the integration
over $U_4(0,\vec{x})$.

We are interested in the thermal partition function in presence of
a given static background field
$\mathbf{A}^{\mathrm{ext}}(\vec{x})$. In the continuum this can be
obtained by splitting the gauge field into the background field
$\mathbf{A}^{\mathrm{ext}}(\vec{x})$ and the fluctuating fields
$\boldsymbol{\eta}(x)$. So that we could write formally for the
thermal partition function
$\mathcal{Z}_T[\mathbf{A}^{\mathrm{ext}}]$:
\begin{equation}
\label{ZetaT} \mathcal{Z}_T[\mathbf{A}^{\mathrm{ext}}]   \equiv
\int \mathcal{D} \boldsymbol{\eta} \, \langle
\mathbf{A}^{\mathrm{ext}}, \boldsymbol{\eta} \left| e^{-\beta_T H}
\mathcal{P} \right| \mathbf{A}^{\mathrm{ext}}, \boldsymbol{\eta}
\rangle \,.
\end{equation}
Actually, to give a meaning to Eq.~(\ref{ZetaT}) one must define
the states $|\mathbf{A}^{\mathrm{ext}}, \boldsymbol{\eta}>$. For
instance, in a perturbative approach, a meaningful definition of
$\mathcal{Z}_T[\mathbf{A}^{\mathrm{ext}}]$ is given by the
background field method. In the non-perturbative lattice approach
the implementation of Eq.~(\ref{ZetaT}) could be obtained as
follows. We write
\begin{equation}
\label{ukbetat}
U_k(\beta_T,\vec{x})=U_k(0,\vec{x})=U^{\text{ext}}_k(\vec{x})
\widetilde{U}_k(\vec{x}) \,,
\end{equation}
where $U^{\text{ext}}_k(\vec{x})$ is related to
$\mathbf{A}^{\mathrm{ext}}(\vec{x})$ by Eq.~(\ref{links}) and the
$\widetilde{U}_k(\vec{x})$'s are the fluctuating links (related to
the fluctuating fields $\boldsymbol{\eta}(x)$) . Thus
Eq.~(\ref{trace}) suggests the following definition
\begin{equation}
\label{zzz} \mathcal{Z}_T[\mathbf{A}^{\mathrm{ext}}] = \int \,
\mathcal{D}\widetilde{U}_k(\vec{x}) \, \,
\mathcal{Z}[U_k^{\text{ext}}(\vec{x}),\widetilde{U}_k(\vec{x})]
\,,
\end{equation}
where we integrate over the fluctuating links
$\widetilde{U}_k(\vec{x})$, while the $U_k^{\text{ext}}(\vec{x})$
links are fixed. Note that in Eq.~(\ref{zzz}) only the spatial
links exiting from sites  belonging to the hyperplane $x_4=0$ are
written as products of the external links
$U^{\text{ext}}_k(\vec{x})$ and the fluctuating links
$\widetilde{U}_k(\vec{x})$. They will be called "frozen links",
while the remainder will be called ``dynamical links''. From the
physical point of view we are considering the gauge system at
finite temperature in interaction with a fixed external background
field. Therefore in the Wilson action $S_W$ we do not include the
contribution of plaquettes built up with only frozen links. The
temporal links $U_4(x_4=0,\vec{x})$ are not constrained and
satisfy the usual periodic boundary conditions. We stress that
p.b.c.'s in temporal direction are crucial to retain the physical
interpretation of the functional
$\mathcal{Z}_T[\mathbf{A}^{\text{ext}}]$ as thermal partition
function.

Now, it is easy to see that in Eq.~(\ref{zzz}) we have
\begin{equation}
\label{ZetaText}
\mathcal{Z} \left[ U_k^{\text{ext}}(\vec{x}),
\widetilde{U}_k(\vec{x}) \right]
= \mathcal{Z} \left[U_k^{\text{ext}}(\vec{x}) \right] \,.
\end{equation}
Indeed, let us consider an arbitrary frozen link
$U^{\text{ext}}_k(\vec{x}) \widetilde{U}_k(\vec{x})$. This link
enters in the Wilson action through the plaquette:
\begin{equation}
\label{plaquette}
P_{k4}(x_4=0,\vec{x}) = \text{Tr} \left\{
U^{\text{ext}}_k(\vec{x})  \widetilde{U}_k(\vec{x})
U_4(0,\vec{x}+\hat{k}) U^\dagger_k(1,\vec{x})
U^\dagger_4(0,\vec{x}) \right\} \,.
\end{equation}
Now we observe that the link $U_4(0,\vec{x}+\hat{k})$ in
Eq.~(\ref{plaquette}) is a dynamical one, i.e. we are integrating
over it. So that the dependence on $\widetilde{U}_k(\vec{x})$ can
be re-absorbed by a change of integration variable. Therefore we
obtain
\begin{equation}
\label{plaqnew}
P_{k4}(x_4=0,\vec{x}) = \text{Tr} \left\{
U^{\text{ext}}_k(\vec{x})
U_4(0,\vec{x}+\hat{k}) U^\dagger_k(1,\vec{x})
U^\dagger_4(0,\vec{x}) \right\} \,.
\end{equation}
It is evident that Eq.~(\ref{plaqnew}) in turns implies
Eq.~(\ref{ZetaText}). Then, we see that in Eq.~(\ref{zzz}) the
integration over the fluctuating links $\widetilde{U}(\vec{x})$
gives an irrelevant multiplicative constant. So that we are led to
define the lattice thermal partition function in presence of a
static background field as
\begin{equation}
\label{ZetaTnew} \mathcal{Z}_T \left[ \mathbf{A}^{\text{ext}}
\right] =
\int_{U_k(\beta_T,\vec{x})=U_k(0,\vec{x})=U^{\text{ext}}_k(\vec{x})}
\mathcal{D}U \, e^{-S_W}   \,.
\end{equation}
Note that the thermal partition function $\mathcal{Z}_T
[\mathbf{A}^{\text{ext}}]$, as defined by Eq.~(\ref{ZetaTnew}), is
invariant for time-independent gauge transformations of the
background field $\mathbf{A}^{\text{ext}}$. On a lattice with
finite spatial extension we impose that the spatial links exiting
from sites belonging to the  spatial boundary of a generic time
slice $x_4$ ($x_4 \ne 0$) are fixed according to the boundary
conditions Eq.~(\ref{coldwall}) used in the Schr\"odinger
functional. Thus we see that, sending physical temperature to
zero, the thermal functional Eq.~(\ref{ZetaTnew}) reduces to the
zero-temperature Schr\"odinger functional Eq.~(\ref{Zetal}) on a
finite lattice.

Let us now introduce our disorder parameter to detect monopole or
vortex condensation. At zero-temperature~\cite{Cea:2000zr}
\begin{equation}
\label{disorder}
\mu = e^{-E_{\text{b.f.}} L_4} = \frac{\mathcal{Z} \left[
\mathbf{A}^{\text{ext}} \right]}{\mathcal{Z}[0]} \,,
\end{equation}
where $\mathcal{Z}[ \mathbf{A}^{\text{ext}}]$ is the lattice
Schr\"odinger functional with monopole or vortex static background
field $\mathbf{A}^{\text{ext}}$. According to the physical
interpretation of the effective action Eq.~(\ref{Gamma})
$E_{\text{b.f.}}$ is the energy to create a monopole or a vortex
in the quantum vacuum. If there is condensation, then
$E_{\text{b.f.}}=0$ and $\mu = 1$.

At finite temperature the disorder parameter turns out to be
related to the monopole or vortex free
energy~\cite{Cea:2000zr,Cea:2000rj}. In particular, the finite
temperature disorder parameter is defined by means of the thermal
partition function in presence of the given static background
field Eq.~(\ref{ZetaTnew}):
\begin{equation}
\label{disorderT}
\mu = e^{-F_{\text{b.f.}}/T_{\text{phys}}} = \frac{\mathcal{Z}_T
\left[ \mathbf{A}^{\text{ext}} \right]} {\mathcal{Z}_T[0]} \,,
\end{equation}
From Eq.~(\ref{disorderT}) it is now clear that $F_{\text{b.f.}}$
is the free energy to create a monopole or a vortex. If there is
condensation, then $F_{\text{b.f.}}=0$ and $\mu = 1$.

As already stated, our disorder parameter is gauge-invariant for
time-independent gauge transformations of the external background
fields, since it has been defined in terms of the Schr\"odinger
functional. Let us stress that gauge invariance for
time-independent gauge transformations of the external background
fields implies that we do not need to do any gauge fixing to
perform the Abelian projection. Indeed, after choosing the Abelian
direction, needed to define the Abelian monopole or vortex fields
through the Abelian projection, due to gauge invariance of
Schr\"odinger functional for transformations of background field,
our results do not depend on the selected Abelian direction,
which, actually, can be varied by a gauge transformation.

The plan of the paper is the following. In Sect.~2 we study the
condensation of vortices and monopoles in the zero temperature
lattice U(1) pure gauge theory. In Sect.~3 we compare Abelian
monopole and vortex condensation for finite temperature SU(2)
lattice gauge theory. Sect.~4 is devoted to finite temperature
SU(3) gauge theory where,  according to the choice of the Abelian
subgroup, two different kinds of Abelian monopoles and vortices
can be defined. Our conclusions are drawn in Sect.~5.
\section{U(1)}
In this Section we study the monopole and vortex condensation in
lattice pure gauge U(1) theory at zero physical temperature. The
disorder parameter~Eq.~(\ref{disorder}) is defined in terms of the
lattice effective action Eq.~(\ref{Gamma}) with a Dirac magnetic
monopole or a magnetic vortex background field.
%
\FIGURE[!ht]{ \label{Fig1}
\includegraphics[clip,width=0.85\textwidth]{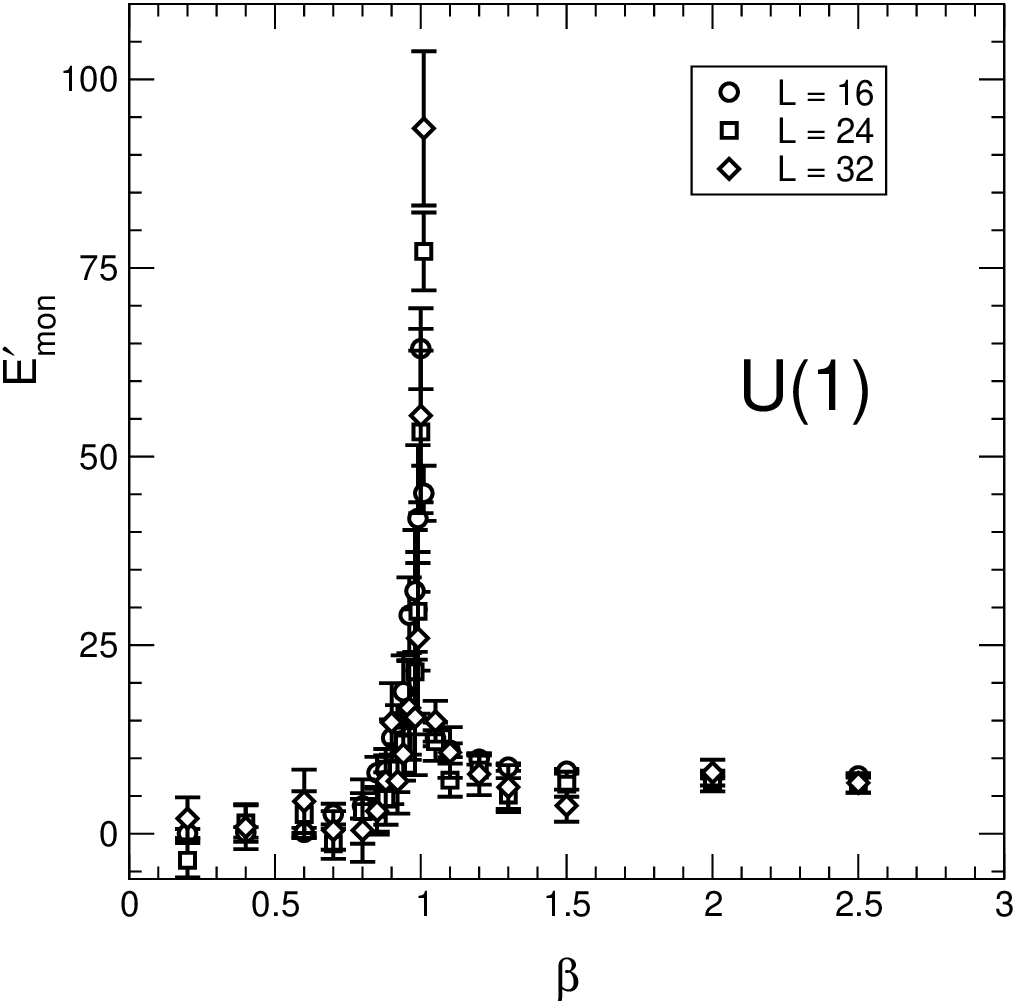}
\caption{The derivative of the energy to create a monopole
($n_{\text{mon}}=1$), Eq.~(\ref{avplaqu1}), versus $\beta$ for
U(1) lattice gauge theory on a $L^4$ lattice (circles refer to
$L=16$, squares to $L=24$, and diamonds to $L=32$).} }

Let us start with the case of the Dirac magnetic monopole
background field. In the continuum the magnetic monopole field
with the Dirac string in the direction $\vec{n}$ is
\begin{equation}
\label{monopu1}
e \vec{b}({\vec{r}}) =  \frac{n_{\mathrm{mon}}}{2} \frac{ \vec{r}
\times \vec{n}}{r(r - \vec{r}\cdot\vec{n})} \,,
\end{equation}
where, according to the Dirac quantization condition,
$n_{\mathrm{mon}}$ is an integer and $e$ is the electric charge
(magnetic charge = $n_{\mathrm{mon}}/2e$). We consider the
gauge-invariant background field action Eq.~(\ref{Gamma}) where
the external background field is given by the lattice version of
the Dirac magnetic monopole field. By choosing $\vec{n}=\hat{x}_3$
we get:
\begin{equation}
\label{monu1links}
\begin{split}
U^{\text{ext}}_{1,2}(\vec{x})  & = \cos [
\theta^{\text{mon}}_{1,2}(\vec{x}) ] + i  \sin [
\theta^{\text{mon}}_{1,2}(\vec{x}) ] \,, \\
U^{\text{ext}}_{3}(\vec{x}) & = {\mathbf 1} \,,
\end{split}
\end{equation}
with
\begin{equation}
\label{monu1theta}
\begin{split}
\theta^{\text{mon}}_1(\vec{x}) & = -\frac{n_{\text{mon}}}{2}
\frac{(x_2-X_2)}{|\vec{x}_{\text{mon}}|}
\frac{1}{|\vec{x}_{\text{mon}}| - (x_3-X_3)} \,, \\
\theta^{\text{mon}}_2(\vec{x}) & = +\frac{n_{\text{mon}}}{2}
\frac{(x_1-X_1)}{|\vec{x}_{\text{mon}}|}
\frac{1}{|\vec{x}_{\text{mon}}| - (x_3-X_3)} \,.
\end{split}
\end{equation}
In Equation~(\ref{monu1theta}) $(X_1,X_2,X_3)$ are the monopole
coordinates and $\vec{x}_{\text{mon}} = (\vec{x} - \vec{X})$. In
the numerical simulations we put the lattice Dirac monopole at the
center of the time slice $x_4=0$. To avoid the singularity due to
the Dirac string we locate the monopole between two neighboring
sites. We have checked that the numerical results are not too
sensitive to the precise position of the magnetic monopole. \\
To avoid the problem of dealing with a partition function we
consider $E^\prime_{\mathrm{mon}} = \partial E_{\mathrm{mon}} /
\partial \beta$. It is easy to see that $E^\prime_{\mathrm{mon}}$
is given by the difference between the average plaquette
$<U_{\mu\nu}>$ obtained in turn from configurations without and
with the monopole field:
\begin{equation}
\label{avplaqu1}
E^\prime_{\text{mon}} = V \left[ <U_{\mu\nu}>_{n_{\text{mon}}=0} -
<U_{\mu\nu}>_{n_{\text{mon}} \ne 0} \right] \,,
\end{equation}
where $V$ is the spatial volume.
%
\FIGURE[!ht]{ \label{Fig2}
\includegraphics[clip,width=0.85\textwidth]{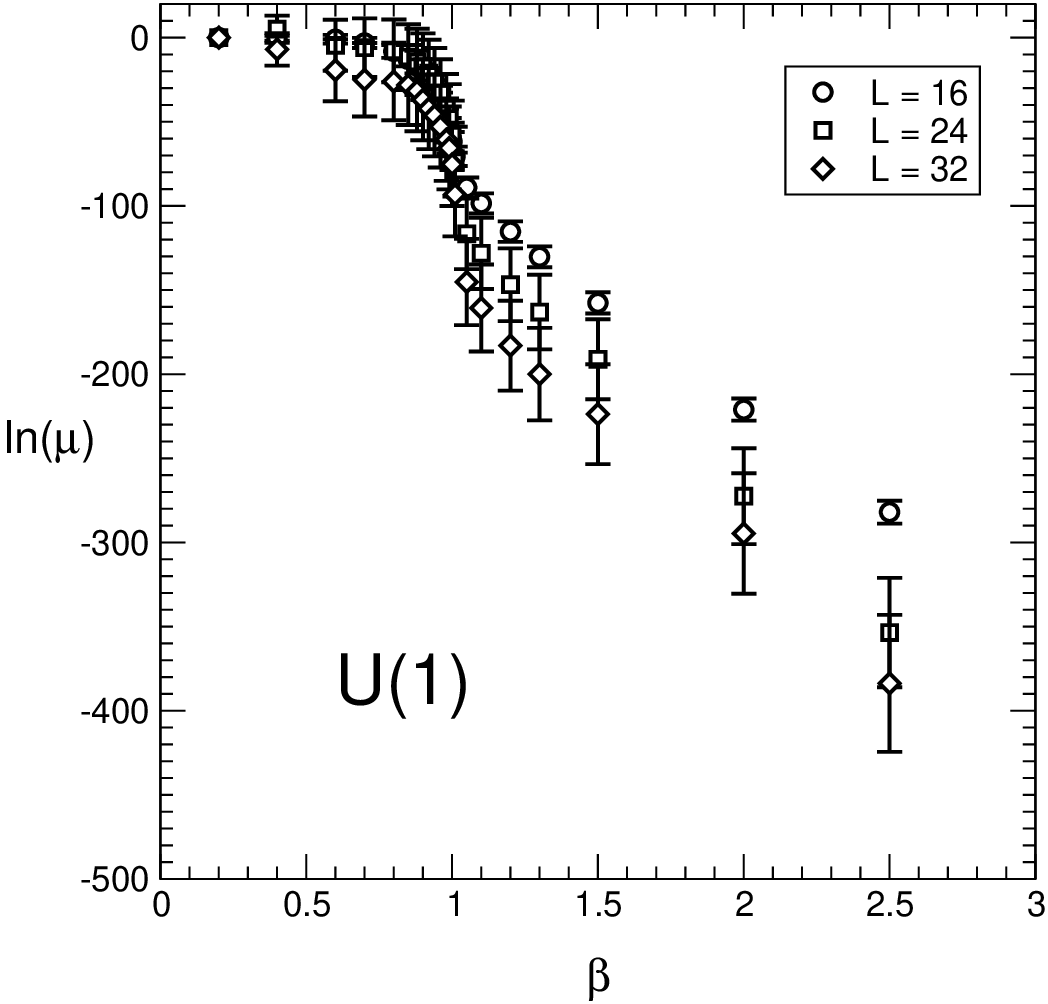}
\caption{The logarithm of the disorder parameter,
Eq.~(\ref{disorder}), versus $\beta$ for U(1) lattice gauge theory
on a $L^4$ lattice. Symbols as in Fig.~1.} }

We performed lattice simulations on $16^4$, $24^4$ and $32^4$
lattices using an APE100 computer. The spatial links belonging to
the time slice $x_4=0$ and to the spatial boundaries of the other
time slices ($x_4\ne0$) are constrained, therefore they are not
updated during Monte Carlo simulation. Therefore we must consider
only the ``dynamical links'' in the computation of
$E^\prime_{\text{mon}}$. This means that the generic plaquette
$U_{\mu\nu}(x)=U_\mu(x)U_\nu(x+\hat{\mu})U^\dagger_\mu(x+\hat{\nu})U^\dagger_\nu(x)$
contributes to Eq.~(\ref{avplaqu1}) if the link $U_\mu(x)$ is a
"dynamical" one (i.e. it is not constrained in the lattice
Schro\"dinger functional integration).
\\
Since we are measuring a local quantity such as average plaquette
$<U_{\mu\nu}>$ , a low statistics (from 1000 up to 5000
configurations) is required in order to get a good estimate of
$E^\prime_{\mathrm{mon}}$. Statistical errors have been estimated
using the jackknife resampling method~\cite{Efron:1982,Shao:1995},
modified to take into account
the statistical correlations between lattice configurations.\\
In Fig.~1 we report our numerical results for
$E^\prime_{\mathrm{mon}}$ versus $\beta$ for three different
lattice sizes.
%
\FIGURE[!ht]{ \label{Fig3}
\includegraphics[clip,width=0.85\textwidth]{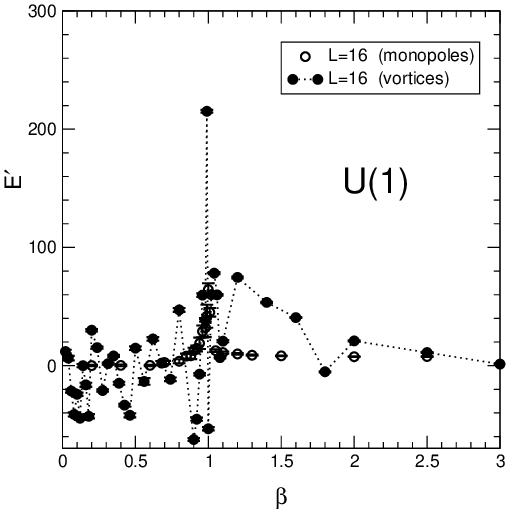}
\caption{The derivative of the energy to create a monopole or a
vortex versus $\beta$ for U(1) lattice gauge theory on a $16^4$
lattice. Open circles refer to a monopole background field
($n_{\text{mon}}=1$) and full circles to a vortex background field
($n_{\text{vort}}=1$) .} }
%
We see that in strong coupling region ($\beta \lesssim 1$),
monopole internal energy derivative is zero, insensitive to the
lattice size. This means that, according to  Eq.~(\ref{disorder}),
the disorder parameter $\mu \simeq 1$. On the other hand, near the
critical coupling $\beta_c \simeq 1$, $E^\prime_{\mathrm{mon}}$
displays a sharp peak which increases by increasing the lattice
volume. In the weak coupling region ($\beta \gg \beta_c$) the
plateau in $E^\prime_{\mathrm{mon}}$ indicates that the monopole
energy tends to the classical monopole action which
behaves linearly  in $\beta$.\\
In order to obtain $\mu$ we perform the numerical integration of
$E^\prime_{\mathrm{mon}}$
\begin{equation}
\label{trapezu1}
E_{{\mathrm{mon}}}  = \int_0^\beta E^\prime_{\mathrm{mon}}
\,d\beta^{\prime}
\end{equation}
In  Fig.~2 we display the logarithm of the disorder parameter
$\mu$ versus $\beta$. In the confined phase we see that $\ln \mu
=0$, so that Eq.~(\ref{disorder}) tells us that the energy
required to create a monopole is zero and therefore monopoles
condense in the confined U(1) vacuum. Correspondingly the disorder
parameter $\mu$ is different from zero in the confined phase.
Moreover numerical data suggest that $\mu \to 0$ when $\beta \to
\beta_c$ in the thermodynamic limit. Note that the different
curves for $\mu$ corresponding to increasing lattice sizes seem to
cross, suggesting a first order phase transition.
However, in order to extract the critical parameters and to determine
the order of transition we need to perform a finite size scaling analysis,
which will be addressed in a future work. 

Let us now consider the case of U(1) magnetic vortices. The
continuum gauge potential for a classical magnetic vortex along
the $x_3$-direction with $n_{\text{vort}}$ units of elementary
flux $\phi=2\pi/e$ is given by
\begin{equation}
\label{u1vort}
\begin{split}
A^{\text{ext}}_1 &= -\frac{n_{\text{vort}}}{e} \frac{x_2}{(x_1)^2
+ (x_2)^2} \,,
\\ A^{\text{ext}}_2 &=  \frac{n_{\text{vort}}}{e} \frac{x_1}{(x_1)^2 + (x_2)^2}
\,, \\ A^{\text{ext}}_3 &=0  \,.
\end{split}
\end{equation}
The corresponding lattice links are:
\begin{equation}
\label{u1links}
\begin{split}
U^{\text{ext}}_{1,2}(\vec{x})  & = \cos
[\theta^{\text{vort}}_{1,2}(\vec{x})] + i \sin
[\theta^{\text{vort}}_{1,2}(\vec{x})] \,,
\\ \theta^{\text{vort}}_{1,2} &= \mp \, n_{\text{vort}} \, \frac{x_{2,1}}{(x_1)^2 + (x_2)^2} \,,
\\U^{\text{ext}}_{3}(\vec{x})  & = \mathbf{1} \,.
\end{split}
\end{equation}
As in the monopole case we evaluated numerically the
$\beta$-derivative, $E^\prime_{\mathrm{vort}}$, of the energy to
create a vortex. In Fig.~3 we compare the monopole and vortex
energy derivative for a $16^4$ lattice . While we clearly see that
in the strong coupling monopoles condense,  the $\beta$-derivative
of the vortex energy displays rapid oscillations whose amplitude
seems to increase near the critical coupling. This means that we
cannot obtain a reliable estimate of $E_{\mathrm{vort}}$ by a
numerical integration of $E^\prime_{\mathrm{vort}}$. However, we
can safely say that our data do not show any signal of vortex
condensation which would imply that
$E^{\prime}_{\text{vort}}=0$ at strong couplings.\\
Thus, we may conclude that in U(1) lattice theory the strong
coupling confined phase is intimately related to magnetic monopole
condensation.
\section{SU(2)}
%
%
%
%
\FIGURE[!ht]{ \label{Fig5}
\includegraphics[clip,width=0.85\textwidth]{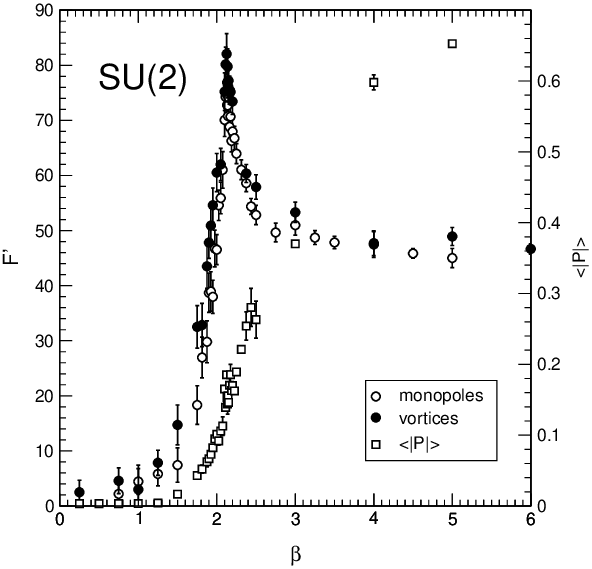}
\caption{The derivative of the free energy,
Eq.~(\ref{derivativesu2}), versus $\beta$ for monopoles (open
circles) and vortices (full circles) for SU(2) on a $24^3 \times
4$ lattice. The absolute value of the Polyakov loop,
Eq.~(\ref{abspolysu2}), is also displayed (open squares). } }
In a previous work~\cite{Cea:2000zr} we studied Abelian magnetic
monopole condensation in finite temperature SU(2) lattice gauge
theory. For SU(2) the maximal Abelian group is an Abelian U(1)
group. In the continuum the Abelian monopole field is given by
\begin{equation}
\label{monop3su2}
g \vec{b}^a({\vec{x}}) = \delta^{a,3} \frac{n_{\mathrm{mon}}}{2}
\frac{ \vec{x} \times \vec{n}}{|\vec{x}|(|\vec{x}| -
\vec{x}\cdot\vec{n})} \,,
\end{equation}
where $\vec{n}$ is the direction of the Dirac string and,
according to the Dirac quantization condition, $n_{\text{mon}}$ is
an integer. The lattice links corresponding to the Abelian
monopole field Eq.~(\ref{monop3su2}) can be readily obtained from
Eq.~(\ref{links}). By choosing $\vec{n}=\hat{x}_3$ we have:
\begin{equation}
\label{su2links}
\begin{split}
U^{\text{ext}}_{1,2}(\vec{x})  & = \cos [
\theta^{\text{mon}}_{1,2}(\vec{x}) ] + i \sigma_3 \sin [
\theta^{\text{mon}}_{1,2}(\vec{x}) ] \,, \\
U^{\text{ext}}_{3}(\vec{x}) & = {\mathbf 1} \,,
\end{split}
\end{equation}
with
\begin{equation}
\label{thetat3su2}
\begin{split}
\theta^{\text{mon}}_1(\vec{x}) & = -\frac{n_{\text{mon}}}{4}
\frac{(x_2-X_2)}{|\vec{x}_{\text{mon}}|}
\frac{1}{|\vec{x}_{\text{mon}}| - (x_3-X_3)} \,, \\
\theta^{\text{mon}}_2(\vec{x}) & = +\frac{n_{\text{mon}}}{4}
\frac{(x_1-X_1)}{|\vec{x}_{\text{mon}}|}
\frac{1}{|\vec{x}_{\text{mon}}| - (x_3-X_3)} \,,
\end{split}
\end{equation}
where  $(X_1,X_2,X_3)$ are the monopole coordinates,
$\vec{x}_{\text{mon}} = (\vec{x} - \vec{X})$ and the $\sigma_a$'s
are the Pauli matrices.

As discussed in Sect.~1, at finite temperature the disorder
parameter, Eq.~(\ref{disorderT}), is defined by means of the
thermal partition function $\mathcal{Z}_T[A^{\text{ext}}]$ in
presence of the Abelian monopole background field
Eq.~(\ref{monop3su2}). Numerical results in Ref.~\cite{Cea:2000zr}
show that the monopole disorder parameter $\mu$ is different from
zero in the confined phase and suggest that it tends to zero when
approaching the critical coupling in the thermodynamic limit. Thus
SU(2) confining vacuum does display the Abelian monopole
condensation in accordance with dual superconductivity hypothesis.

Now we want discuss what happens if we consider an Abelian vortex
background field. In SU(2) gauge theory, Abelian vortex field on
the lattice is given by
\begin{equation}
\label{su2vortlinks}
\begin{split}
U^{\text{ext}}_{1,2}(\vec{x})  & = \cos [
\theta^{\text{vort}}_{1,2}(\vec{x}) ] + i \sigma_3 \sin [
\theta^{\text{vort}}_{1,2}(\vec{x}) ] \,, \\
U^{\text{ext}}_{3}(\vec{x}) & = {\mathbf 1} \,,\\
\theta^{\text{vort}}_{1,2}(\vec{x}) & = \mp
\frac{n_{\text{vort}}}{2} \frac{x_{2,1}}{(x_1)^2+(x_2)^2}  \,.
\end{split}
\end{equation}
Again, the derivative of the free energy required to create a
vortex
\begin{equation}
\label{derivativesu2}
F^\prime_{\text{vort}} = \frac{\partial}{\partial \beta}
F_{\text{vort}} \,,
\end{equation}
can be easily evaluated as the difference between the average
plaquette $<U_{\mu\nu}>$ without the vortex background field (i.e.
$n_{\text{vort}}=0$) and with the vortex background field
($n_{\text{vort}} \ne 0$)
\begin{equation}
\label{avplaqsu2}
F^\prime_{\text{vort}} = V \left[ <U_{\mu\nu}>_{n_{\text{vort}}=0}
- <U_{\mu\nu}>_{n_{\text{vort}} \ne 0} \right] \,,
\end{equation}
where $V$ is the spatial volume.  In Eq.~(\ref{avplaqsu2}) we
include only the contributions due to the dynamical links (see
discussion after Eq.~(\ref{avplaqu1})).
%
\FIGURE[!ht]{
\label{Fig6}
\includegraphics[clip,width=0.85\textwidth]{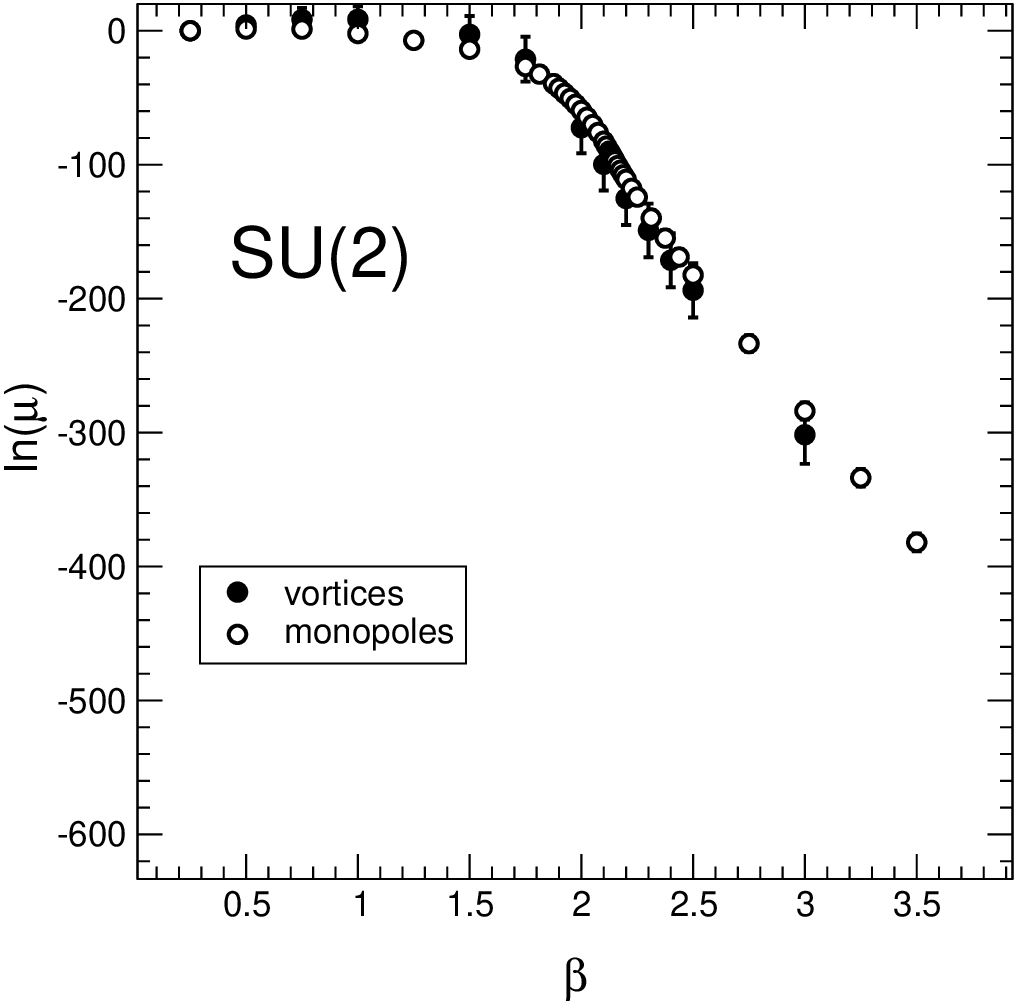}
\caption{The logarithm of the disorder parameter, Eq.~(\ref{disorderT}), versus $\beta$
for vortices (full circles) and monopoles (open circles). Data refer to SU(2) gauge
theory on a $24^3 \times 4$ lattice.}
}
%
A low statistics (from 2000 up to 10000 configurations) is
required in order to get a good estimate of the derivative of the
free energy. In Fig.~4 we display the derivative of the vortex
free energy versus $\beta$ for $n_{\text{vort}}=10$ on lattices
with $L_t =4$ and $L_s = 24$. We see that $F^\prime_{\text{vort}}$
vanishes at strong coupling and displays a rather sharp peak near
$\beta \backsimeq 2.2$. We expect that this peak corresponds to
the finite temperature deconfinement transition. In Fig.~4 we also
display the absolute value of the Polyakov loop in time direction
\begin{equation}
\label{abspolysu2}
P = \frac{1}{V} \sum_{\vec{x}} \frac{1}{2} {\text{Tr}}
\prod_{x_4=1}^{L_t} U_4(x_4,\vec{x})
\end{equation}
and, indeed, we can see that the peak in $F^\prime_{\text{vort}}$
corresponds to the rise of the Polyakov loop.\\
In weak coupling region the plateau in $F^\prime_{\text{vort}}$
indicates that vortex free energy tends to the classical vortex
action which behaves linearly in $\beta$.
%
%
\FIGURE[!ht]{
\label{Fig7}
\includegraphics[clip,width=0.85\textwidth]{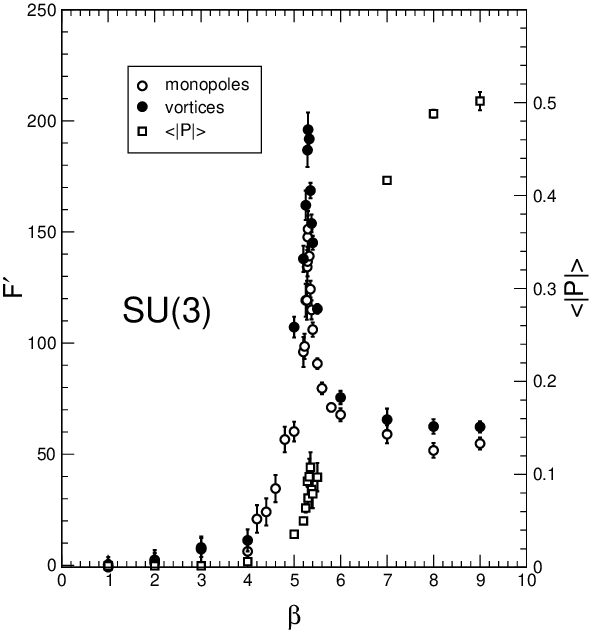}
\caption{The derivative of the free energy versus $\beta$ for monopoles (open
circles) and vortices (full circles). The absolute value of the Polyakov loop,
Eq.~(\ref{abspolysu3}), is also displayed (open squares).
Data refer to SU(3) gauge theory on a $32^3 \times 4$ lattice.}
}
In Fig.~4 we display for comparison also the derivative of the
Abelian monopole free energy for $n_{\text{mon}}=10$. It turns out
that monopoles and vortices data agree within statistical errors.
To appreciate better this last point we plotted in Fig.~5 the
logarithm of the disorder parameter Eq.~(\ref{disorderT}) for
monopoles and vortices respectively. Fig.~5 shows clearly that
monopoles and vortices agree quite perfectly. Our results strongly
suggest that also Abelian vortices could play a role in the
dynamics of confinement.
\FIGURE[!ht]{
\label{Fig8}
\includegraphics[clip,width=0.85\textwidth]{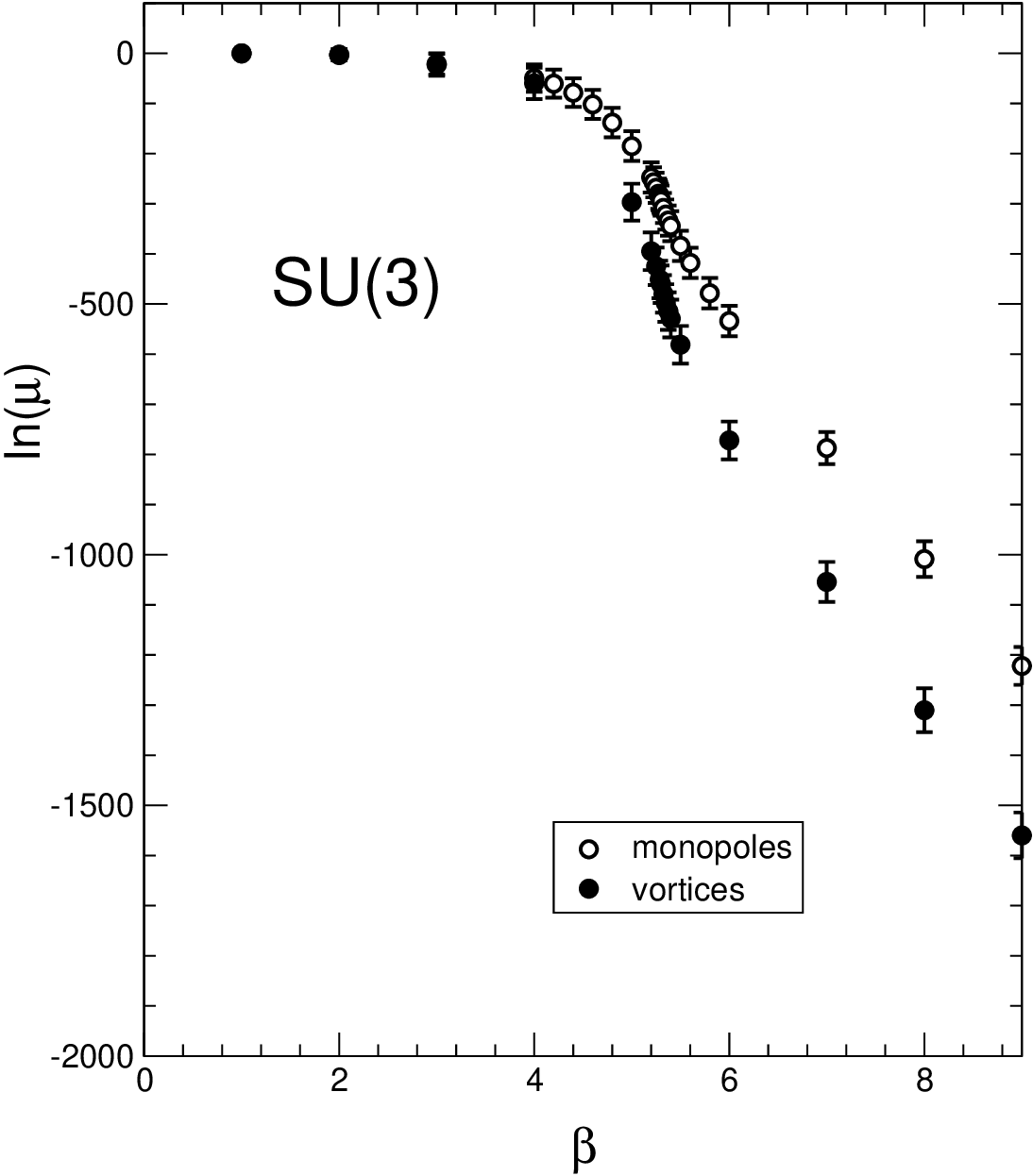}
\caption{The logarithm of the disorder parameter, Eq.~(\ref{disorderT}), versus $\beta$
for $T_8$ monopoles (open circles) and $T_8$ vortices (full circles).
Data refer to SU(3) gauge theory on a $32^3 \times 4$ lattice.}
}
%
\FIGURE[!ht]{
\label{Fig9}
\includegraphics[clip,width=0.85\textwidth]{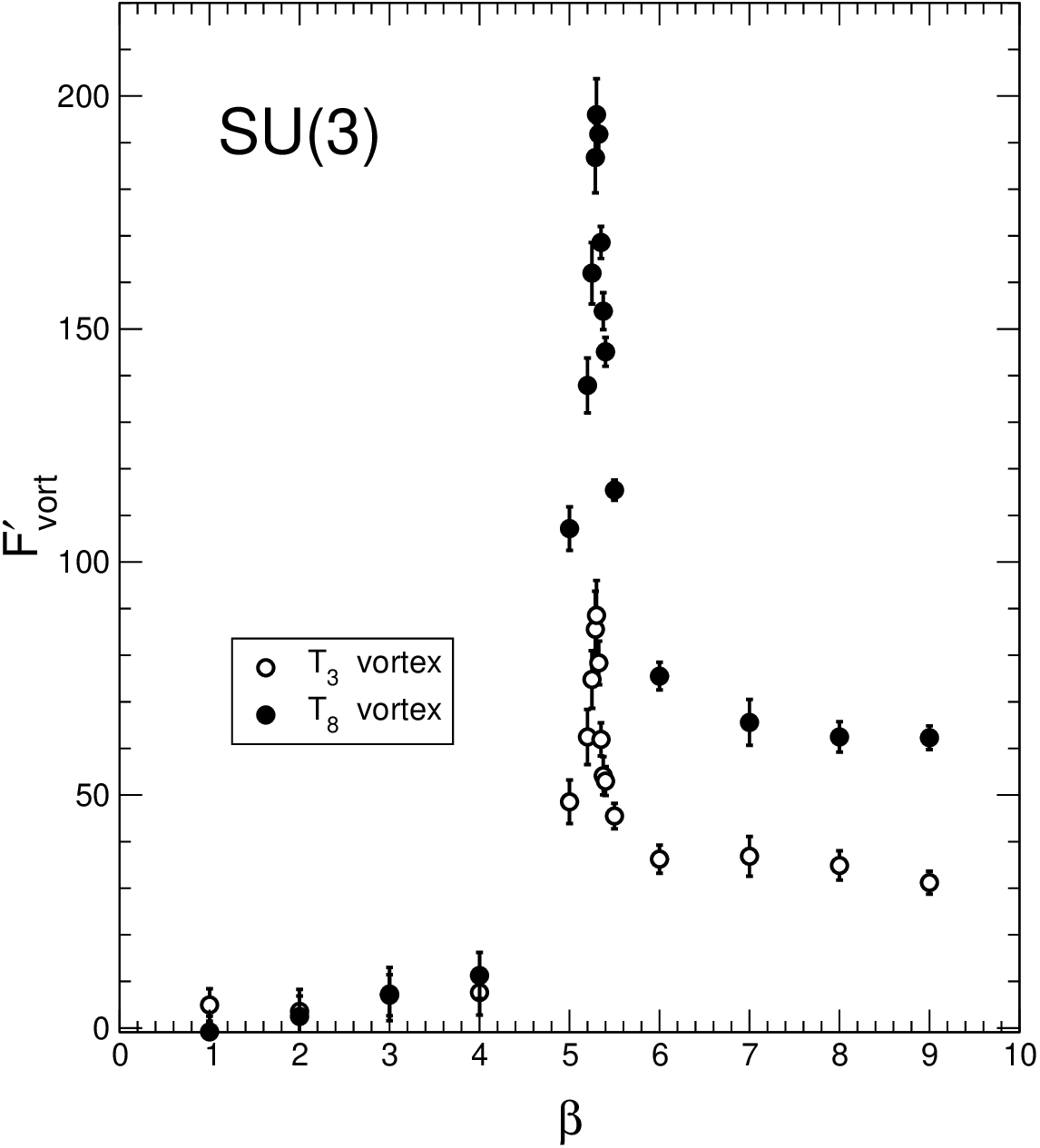}
\caption{The free energy derivative for $T_3$ vortices (open circles)
and $T_8$ vortices (full circles) versus $\beta$.
Data refer to SU(3) gauge theory on a $32^3 \times 4$ lattice.}
}
%
\FIGURE[!ht]{
\label{Fig10}
\includegraphics[clip,width=0.85\textwidth]{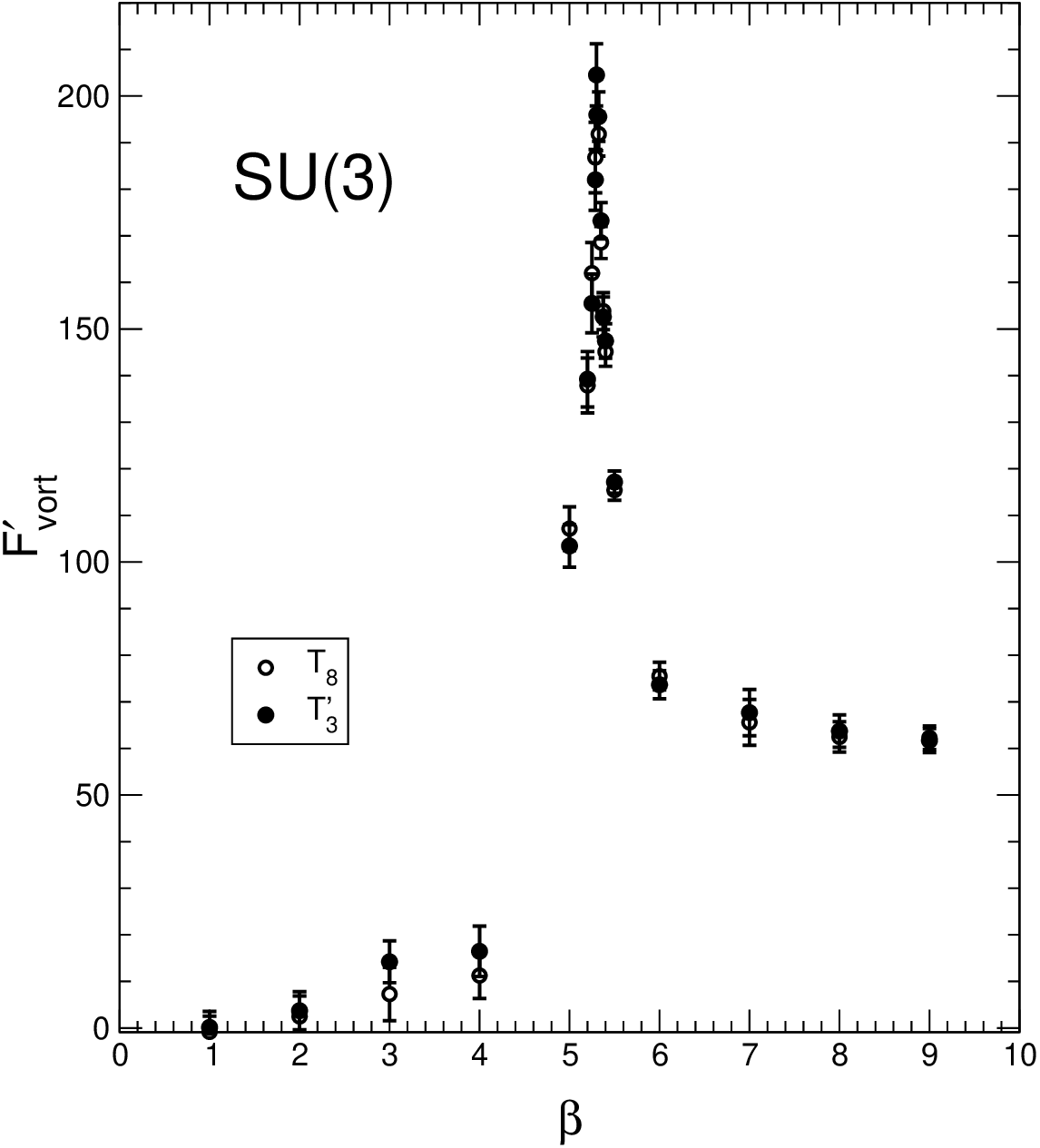}
\caption{The free energy derivative for $T_8$ vortices (open
circles) and $T^{\prime}_{3}$ vortices (full circles) versus
$\beta$. Data refer to SU(3) gauge theory on a $32^3 \times 4$
lattice.} }
\FIGURE[!ht]{
\label{Fig11}
\includegraphics[clip,width=0.85\textwidth]{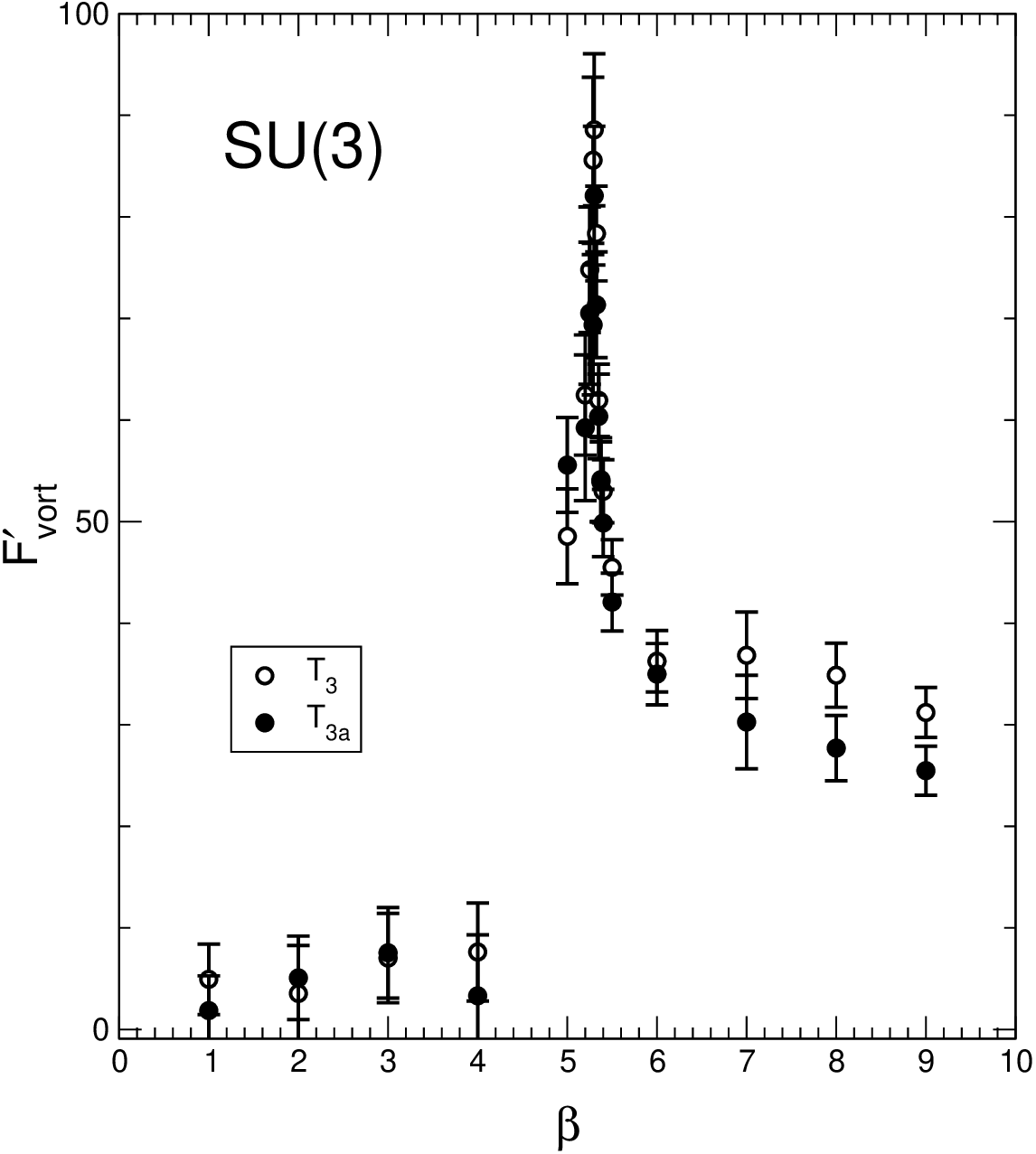}
\caption{The free energy derivative for $T_3$ vortices (open circles)
and $T_{3a}$ vortices (full circles) versus $\beta$.
Data refer to SU(3) gauge theory on a $32^3 \times 4$ lattice.}
}
%
\FIGURE[!ht]{
\label{Fig12}
\includegraphics[clip,width=0.85\textwidth]{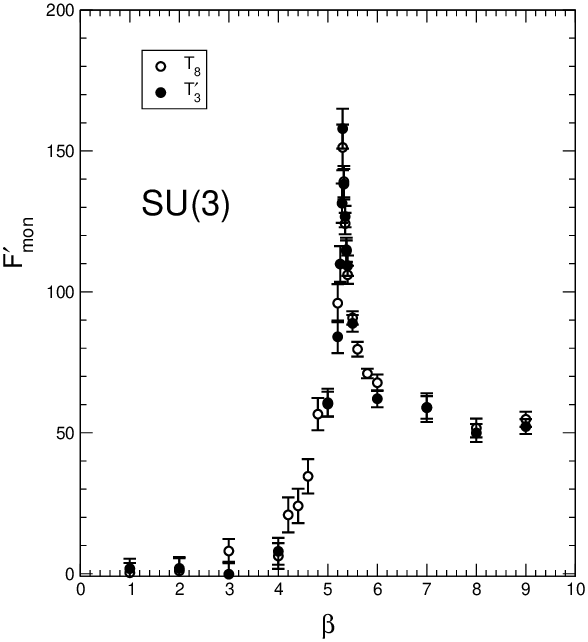}
\caption{The free energy derivative for $T_8$ monopoles (open
circles) and $T^{\prime}_{3}$ monopoles (full circles) versus
$\beta$. Data refer to SU(3) gauge theory on a $32^3 \times 4$
lattice.} }
%
%
%
\section{SU(3)}
For SU(3) gauge theory the maximal Abelian group is
U(1)$\times$U(1), therefore we may introduce two independent types
of Abelian monopoles or Abelian vortices.

Let us consider the Abelian monopole field. The first type of
Abelian monopole field is derived considering the $\lambda_3$
diagonal generator, we name it $T_3$ Abelian monopole (following
Ref.~\cite{Cea:2000zr}). On the lattice it is given by
\begin{equation}
\label{t3linkssu3}
\begin{split}
U_{1,2}^{\text{ext}}(\vec{x}) & =
\begin{bmatrix}
e^{i \theta^{\text{mon}}_{1,2}(\vec{x})} & 0 & 0 \\ 0 &  e^{- i
\theta^{\text{mon}}_{1,2}(\vec{x})} & 0 \\ 0 & 0 & 1
\end{bmatrix}
\,  \\ U^{\text{ext}}_{3}(\vec{x}) & = {\mathbf 1} \,,
\end{split}
\end{equation}
with $\theta^{\text{mon}}_{1,2}(\vec{x})$ defined in
Eq.~(\ref{thetat3su2}). \\
The second type of independent Abelian monopole can be obtained by
considering the diagonal generator $\lambda_8$.
In this case we have the  $T_8$ Abelian monopole:
\begin{equation}
\label{t8linkssu3m}
\begin{split}
U_{1,2}^{\text{ext}}(\vec{x}) & =
\begin{bmatrix}
e^{i \theta^{\text{mon}}_{1,2}(\vec{x})} & 0 & 0 \\ 0 &  e^{i
\theta^{\text{mon}}_{1,2}(\vec{x})} & 0 \\ 0 & 0 & e^{- 2 i
\theta^{\text{mon}}_{1,2}(\vec{x})}
\end{bmatrix}
\,  \\ U^{\text{ext}}_{3}(\vec{x}) & = {\mathbf 1} \;  \, ,
\end{split}
\end{equation}
with
\begin{equation}
\label{thetat8su3m}
\begin{split}
\theta^{\text{mon}}_1(\vec{x}) & = \frac{1}{\sqrt{3}} \left[
 -\frac{n_{\text{mon}}}{4}
\frac{(x_2-X_2)}{|\vec{x}_{\text{mon}}|}
\frac{1}{|\vec{x}_{\text{mon}}| - (x_3-X_3)} \right] \,, \\
\theta^{\text{mon}}_2(\vec{x}) & =  \frac{1}{\sqrt{3}} \left[
+\frac{n_{\text{mon}}}{4} \frac{(x_1-X_1)}{|\vec{x}_{\text{mon}}|}
\frac{1}{|\vec{x}_{\text{mon}}| - (x_3-X_3)} \right] \,.
\end{split}
\end{equation}
Analogously, we have the $T_3$ Abelian vortex:
\begin{equation}
\label{t3linkssu3v}
\begin{split}
U_{1,2}^{\text{ext}}(\vec{x}) &=
\begin{bmatrix}
e^{i \theta^{\text{vort}}_{1,2}(\vec{x})} & 0 & 0 \\ 0 &  e^{- i
\theta^{\text{vort}}_{1,2}(\vec{x})} & 0 \\ 0 & 0 & 1
\end{bmatrix}
\,,\\[0.3cm]
 U_3^{\text{ext}}(\vec{x}) &= {\mathbf{1}} \\[0.3cm]
\theta^{\text{vort}}_{1,2} &= \mp  \frac{n_{\text{vort}}}{2}
\frac{x_{2,1}}{(x_1)^2+(x_2)^2} \; \; .
\end{split}
\end{equation}
and the  $T_8$ Abelian vortex:
\begin{equation}
\label{t8linkssu3v}
\begin{split}
U_{1,2}^{\text{ext}}(\vec{x}) &=
\begin{bmatrix}
e^{i \theta^{\text{vort}}_{1,2}(\vec{x})} & 0 & 0 \\ 0 &  e^{i
\theta^{\text{vort}}_{1,2}(\vec{x})} & 0 \\ 0 & 0 & e^{- 2 i
\theta^{\text{vort}}_{1,2}(\vec{x})}
\end{bmatrix}
\,,\\[0.3cm]
 U_3^{\text{ext}}(\vec{x}) &= {\mathbf{1}} \\[0.3cm]
\theta^{\text{vort}}_{1,2} &= \mp \frac{1}{\sqrt{3}}
\frac{n_{\text{vort}}}{2} \frac{x_{2,1}}{(x_1)^2+(x_2)^2} \; \; ,
\end{split}
\end{equation}
Other Abelian monopoles and vortices can be generated by
considering linear combinations of $\lambda_3$ and $\lambda_8$
generators. In particular we have also considered the following
two linear combinations of $\lambda_3/2$ and $\lambda_8/2$
\begin{equation}
\label{T3a} T_{3a} = -\frac{1}{2} \frac{\lambda_3}{2} +
\frac{\sqrt{3}}{2} \frac{\lambda_8}{2} =
\begin{bmatrix}
0 & 0 & 0 \\ 0 &  \frac{1}{2} & 0 \\ 0 & 0 & -\frac{1}{2}
\end{bmatrix}
\,,
\end{equation}
and
\begin{equation}
\label{T3'}
T^{\prime}_{3} = \frac{\sqrt{3}}{2} \frac{\lambda_3}{2} +
\frac{1}{2} \frac{\lambda_8}{2} =
\begin{bmatrix}
\frac{1}{\sqrt{3}}  & 0 & 0\\
0 &  -\frac{1}{2 \sqrt{3}} & 0 \\
0 & 0 &  -\frac{1}{2 \sqrt{3}}
\end{bmatrix}
\,.
\end{equation}
The linear combinations given in Eq.~(\ref{T3a}) and in Eq.~(\ref{T3'})
have been also considered in Ref.~\cite{DiGiacomo:1999fa,DiGiacomo:1999fb}
and in Ref.~\cite{deForcrand:2000pg} respectively.

For reader convenience, let us summarize the main results of our
previous investigation of Abelian monopole
condensation~\cite{Cea:2000zr} in SU(3). In Ref.~\cite{Cea:2000zr}
we found that there is condensation of Abelian monopoles in the
non-perturbative vacuum and that SU(3) vacuum reacts moderately
strongly in the case of the $T_8$ Abelian monopole (in particular
the peak value of $F^\prime_{\text{mon}}$ for $T_8$ Abelian
monopoles is about two times higher than the corresponding peak
for $T_3$ Abelian monopoles).

In the present paper we compare $T_8$ Abelian monopoles and
vortices. In Fig.~6 the free energy derivative for monopoles (with
$n_{\text{mon}}=10$) and vortices (with $n_{\text{vort}}=10$) is
displayed versus $\beta$ for a lattice with $L_s=32$ and $L_t =4$.
We also display the absolute value of the Polyakov loop in the
time direction
\begin{equation}
\label{abspolysu3}
P = \frac{1}{V} \sum_{\vec{x}} \frac{1}{3} {\text{Tr}}
\prod_{x_4=1}^{L_t} U_4(x_4,\vec{x}) \,.
\end{equation}
As can be argued from Fig.~6, $F^\prime_{\text{vort}}$ behaves
like $F^\prime_{\text{mon}}$. Indeed, the free energy derivatives
are zero within errors in the strong coupling region and display a
sharp peak in correspondence of the rise of the Polyakov loop. In
the weak coupling region the free energy derivatives are almost
constant. The values of the plateau correspond to the beta
derivative of the lattice classical action. Remarkably Fig.~6
shows that in the peak region the $T_8$ Abelian vortex displays  a
signal higher than the Abelian monopole. This is better
appreciated if we look at the disorder parameter
Eq.~(\ref{disorderT}) (see Fig.~7), where $\ln \mu$ for Abelian
vortices has a sizeable faster decrease. In the study of Abelian
monopole condensation we found~\cite{Cea:2000zr} that the color
direction $\hat{8}$ is slightly preferred with respect to the
direction $\hat{3}$ in the color space. It is worthwhile to see if
this result holds also for Abelian vortex condensation. In Fig.~8
we compare the free energy derivative for the $T_3$ and $T_8$
Abelian vortices  for the lattice with $L_t =4$ and $L_s = 32$. As
expected  the $T_8$ Abelian vortex displays, in the peak region, a
signal about a factor two higher.

Finally in Fig.~9 and in Fig.~10 we confront  the $T_3$, $T_{3a}$,
$T_8$ and $T^{\prime}_{3}$ Abelian vortices. We find that $T_3$
and $T_8$ agree within statistical errors in the whole range of
$\beta$ with $T_{3a}$ and $T^{\prime}_{3}$ respectively. For the
sake of completeness, in Fig.~11 we compare $T^{\prime}_{3}$ and
$T_8$ monopoles, showing that they agree within statistical
errors.

We may conclude that, even for SU(3) gauge theory, our results
strongly suggest that Abelian vortices play a role in the dynamics
of confinement.
\section{Conclusions}
We investigated Abelian monopole and Abelian vortex condensation
in U(1), SU(2) and SU(3) lattice gauge theories.
\\
For U(1) pure lattice gauge theory we found that, in the confined
phase, the vacuum can be interpreted as a coherent condensate of
magnetic monopoles. On the other hand, we do not find convincing
evidence of condensation of vortices.
\\
For non Abelian SU(2) and SU(3) lattice gauge theories at finite
temperature, by means of a lattice thermal partition functional,
we introduced a disorder parameter for detecting Abelian monopole
and Abelian vortex condensation in the confined phase. The
disorder parameter is defined by means of a lattice thermal
partition functional and is  invariant for gauge transformations
of the external background (monopole or vortex) field.

Our numerical results suggest that the disorder parameter for both
Abelian monopoles and Abelian vortices  is different from zero in
the confined phase and tends to zero when approaching the critical
coupling in the thermodynamic limit. Therefore in SU(2) and SU(3)
Abelian vortices could play a role in the dynamics of
confinement.\\
In particular for SU(2) gauge theory there is also a quantitative
agreement between the measured values of free energies for
monopoles and vortices. \\
On the other hand, remarkably, in SU(3) gauge theory it turns out
that the Abelian vortex displays  a signal higher than the Abelian
monopole. Moreover, for the Abelian vortices we find that the non
perturbative vacuum reacts moderately strongly to $T_8$ vortices
with respect to $T_3$ vortices. This last point is in accordance
with our finding in the study of SU(3) Abelian monopole
condensation~\cite{Cea:2000zr}.

In conclusion it is worthwhile to observe that our results point
to a different mechanism of confinement for U(1) lattice gauge
theory with respect to SU(2) and SU(3) gauge theories. Indeed, in
the U(1) Abelian case we find that the confining vacuum behaves as
a coherent condensate of Dirac magnetic monopoles. In SU(2) and
SU(3) it seems that there is condensation of Abelian magnetic
monopoles and Abelian vortices. So that in SU(2) and SU(3) gauge
theories one could look at the confining vacuum as a coherent
Abelian magnetic condensate. Even more, for the SU(3) theory, it
turns out that gauge field configurations leading to Abelian
magnetic flux tubes, multiple of the elementary flux, seem to be
favorite with respect to Abelian magnetic monopoles.

We would like to remark that it is important to perform a 
finite size scaling analysis both for SU(2) and SU(3).
Indeed a finite size scaling analysis will allow us to
determine the critical behavior of the disorder parameter in 
the thermodynamic limit.

We stress finally, that it should be interesting to extend our
method to study center vortex condensation in SU(2) and SU(3)
lattice gauge theories. First results in this direction have been
reported in Ref.~\cite{Cea:2001nm}.

%
%
%

\begin{thebibliography}{10}

\bibitem{tHooft:1976eps}
G.~'t~Hooft, {\it The confinement phenomenon in quantum field theory},  in {\em
  High Energy Physics, EPS International Conference, Palermo, 1975}.

\bibitem{Mandelstam:1974pi}
S.~Mandelstam, {\it Vortices and quark confinement in non {Abelian} gauge
  theories},  {\em Phys. Rept.} {\bf 23} (1976) 245.

\bibitem{DiGiacomo:1994jy}
A.~{Di Giacomo}, {\it Mechanisms of color confinement},  {\em Acta Phys.
  Polon.} {\bf B25} (1994) 215--226.

\bibitem{DiGiacomo:1999fa}
A.~{Di Giacomo}, B.~Lucini, L.~Montesi, and G.~Paffuti, {\it Colour confinement
  and dual superconductivity of the vacuum. i},  {\em Phys. Rev.} {\bf D61}
  (2000) 034503, [\href{http://xxx.lanl.gov/abs/hep-lat/9906024}{{\tt
  hep-lat/9906024}}].

\bibitem{DiGiacomo:1999fb}
A.~{Di Giacomo}, B.~Lucini, L.~Montesi, and G.~Paffuti, {\it Colour confinement
  and dual superconductivity of the vacuum. ii},  {\em Phys. Rev.} {\bf D61}
  (2000) 034504, [\href{http://xxx.lanl.gov/abs/hep-lat/9906025}{{\tt
  hep-lat/9906025}}].

\bibitem{Carmona:2001ja}
J.~M. Carmona, M.~D'Elia, A.~{Di Giacomo}, B.~Lucini, and G.~Paffuti, {\it
  Color confinement and dual superconductivity of the vacuum. iii},
  \href{http://xxx.lanl.gov/abs/hep-lat/0103005}{{\tt hep-lat/0103005}}.

\bibitem{Cea:2000zr}
P.~Cea and L.~Cosmai, {\it Gauge invariant study of the monopole condensation
  in non {Abelian} lattice gauge theories},  {\em Phys. Rev.} {\bf D62} (2000)
  094510, [\href{http://xxx.lanl.gov/abs/hep-lat/0006007}{{\tt
  hep-lat/0006007}}].

\bibitem{Shiba:1995db}
H.~Shiba and T.~Suzuki, {\it Monopole action and condensation in {SU(2)}
  {QCD}},  {\em Phys. Lett.} {\bf B351} (1995) 519--527,
  [\href{http://xxx.lanl.gov/abs/hep-lat/9408004}{{\tt hep-lat/9408004}}].

\bibitem{Arasaki:1997sm}
N.~Arasaki, S.~Ejiri, S.-i. Kitahara, Y.~Matsubara, and T.~Suzuki, {\it
  Monopole action and monopole condensation in {SU(3)} lattice {QCD}},  {\em
  Phys. Lett.} {\bf B395} (1997) 275--282,
  [\href{http://xxx.lanl.gov/abs/hep-lat/9608129}{{\tt hep-lat/9608129}}].

\bibitem{Nakamura:1997sw}
N.~Nakamura {\em et.~al.}, {\it Disorder parameter of confinement},  {\em Nucl.
  Phys. Proc. Suppl.} {\bf 53} (1997) 512--514,
  [\href{http://xxx.lanl.gov/abs/hep-lat/9608004}{{\tt hep-lat/9608004}}].

\bibitem{Chernodub:1997ps}
M.~N. Chernodub, M.~I. Polikarpov, and A.~I. Veselov, {\it Effective constraint
  potential for {Abelian} monopole in {SU(2)} lattice gauge theory},  {\em
  Phys. Lett.} {\bf B399} (1997) 267--273,
  [\href{http://xxx.lanl.gov/abs/hep-lat/9610007}{{\tt hep-lat/9610007}}].

\bibitem{Jersak:1999nv}
J.~Jersak, T.~Neuhaus, and H.~Pfeiffer, {\it Scaling analysis of the magnetic
  monopole mass and condensate in the pure {U(1)} lattice gauge theory},  {\em
  Phys. Rev.} {\bf D60} (1999) 054502,
  [\href{http://xxx.lanl.gov/abs/hep-lat/9903034}{{\tt hep-lat/9903034}}].

\bibitem{Hoelbling:2000su}
C.~Hoelbling, C.~Rebbi, and V.~A. Rubakov, {\it Free energy of an {SU(2)}
  monopole-antimonopole pair},  {\em Phys. Rev.} {\bf D63} (2001) 034506,
  [\href{http://xxx.lanl.gov/abs/hep-lat/0003010}{{\tt hep-lat/0003010}}].

\bibitem{Faber:1999gu}
M.~Faber, J.~Greensite, S.~Olejnik, and D.~Yamada, {\it The vortex-finding
  property of maximal center (and other) gauges},  {\em JHEP} {\bf 12} (1999)
  012, [\href{http://xxx.lanl.gov/abs/hep-lat/9910033}{{\tt hep-lat/9910033}}].

\bibitem{Langfeld:1998cz}
K.~Langfeld, O.~Tennert, M.~Engelhardt, and H.~Reinhardt, {\it Center vortices
  of {Yang-Mills} theory at finite temperatures},  {\em Phys. Lett.} {\bf B452}
  (1999) 301, [\href{http://xxx.lanl.gov/abs/hep-lat/9805002}{{\tt
  hep-lat/9805002}}].

\bibitem{deForcrand:1999ms}
P.~de~Forcrand and M.~D'Elia, {\it On the relevance of center vortices to
  {QCD}},  {\em Phys. Rev. Lett.} {\bf 82} (1999) 4582--4585,
  [\href{http://xxx.lanl.gov/abs/hep-lat/9901020}{{\tt hep-lat/9901020}}].

\bibitem{Bertle:1999vu}
R.~Bertle, M.~Faber, J.~Greensite, and S.~Olejnik, {\it The structure of
  projected center vortices at zero and finite temperature},  {\em Nucl. Phys.
  Proc. Suppl.} {\bf 83} (2000) 425--427,
  [\href{http://xxx.lanl.gov/abs/hep-lat/9909002}{{\tt hep-lat/9909002}}].

\bibitem{Ambjorn:1999ym}
J.~Ambjorn, J.~Giedt, and J.~Greensite, {\it Vortex structure vs. monopole
  dominance in {Abelian} projected gauge theory},  {\em JHEP} {\bf 02} (2000)
  033, [\href{http://xxx.lanl.gov/abs/hep-lat/9907021}{{\tt hep-lat/9907021}}].

\bibitem{Montero:1999gq}
A.~Montero, {\it {SU(3)} vortex-like configurations in the maximal center
  gauge},  {\em Nucl. Phys. Proc. Suppl.} {\bf 83} (2000) 518--520,
  [\href{http://xxx.lanl.gov/abs/hep-lat/9907024}{{\tt hep-lat/9907024}}].

\bibitem{Jahn:1999ab}
O.~Jahn, F.~Lenz, J.~W. Negele, and M.~Thies, {\it Center vortices, instantons,
  and confinement},  {\em Nucl. Phys. Proc. Suppl.} {\bf 83} (2000) 524--526,
  [\href{http://xxx.lanl.gov/abs/hep-lat/9909062}{{\tt hep-lat/9909062}}].

\bibitem{Stack:2000zx}
J.~D. Stack and W.~Tucker, {\it On the distribution of vortex sizes and the
  long range potential},  {\em Nucl. Phys. Proc. Suppl.} {\bf 83} (2000)
  539--540.

\bibitem{Bakker:1999qh}
B.~L.~G. Bakker, A.~I. Veselov, and M.~A. Zubkov, {\it Central dominance and
  the confinement mechanism in gluodynamics},  {\em Phys. Lett.} {\bf B471}
  (1999) 214--219, [\href{http://xxx.lanl.gov/abs/hep-lat/9902010}{{\tt
  hep-lat/9902010}}].

\bibitem{Alexandru:2000ir}
A.~Alexandru and R.~W. Haymaker, {\it Vortices in {SO(3)} {$\times$} {Z(2)}
  simulations},  {\em Nucl. Phys. Proc. Suppl.} {\bf 94} (2001) 475--477,
  [\href{http://xxx.lanl.gov/abs/hep-lat/0009012}{{\tt hep-lat/0009012}}].

\bibitem{DelDebbio:2000cx}
L.~{Del Debbio}, A.~{Di Giacomo}, and B.~Lucini, {\it Vortices, monopoles and
  confinement},  {\em Nucl. Phys.} {\bf B594} (2001) 287--300,
  [\href{http://xxx.lanl.gov/abs/hep-lat/0006028}{{\tt hep-lat/0006028}}].

\bibitem{deForcrand:2000pg}
P.~de~Forcrand and M.~Pepe, {\it Center vortices and monopoles without lattice
  {Gribov} copies},  {\em Nucl. Phys.} {\bf B598} (2001) 557--577,
  [\href{http://xxx.lanl.gov/abs/hep-lat/0008016}{{\tt hep-lat/0008016}}].

\bibitem{Kovacs:2000sy}
T.~G. Kovacs and E.~T. Tomboulis, {\it Computation of the vortex free energy in
  {SU(2)} gauge theory},  {\em Phys. Rev. Lett.} {\bf 85} (2000) 704--707,
  [\href{http://xxx.lanl.gov/abs/hep-lat/0002004}{{\tt hep-lat/0002004}}].

\bibitem{Korthals-Altes:2000gs}
C.~Korthals-Altes and A.~Kovner, {\it Magnetic {Z(N)} symmetry in hot {QCD} and
  the spatial {Wilson} loop},  {\em Phys. Rev.} {\bf D62} (2000) 096008,
  [\href{http://xxx.lanl.gov/abs/hep-ph/0004052}{{\tt hep-ph/0004052}}].

\bibitem{'tHooft:1978hy}
G.~'t~Hooft, {\it On the phase transition towards permanent quark confinement},
   {\em Nucl. Phys.} {\bf B138} (1978) 1.

\bibitem{Cornwall:1979hz}
J.~M. Cornwall, {\it Quark confinement and vortices in massive gauge invariant
  {QCD}},  {\em Nucl. Phys.} {\bf B157} (1979) 392.

\bibitem{Cornwall:1998ds}
J.~M. Cornwall, {\it Center vortices and confinement vs. screening},  {\em
  Phys. Rev.} {\bf D57} (1998) 7589--7600,
  [\href{http://xxx.lanl.gov/abs/hep-th/9712248}{{\tt hep-th/9712248}}].

\bibitem{Yaffe:1980iq}
L.~G. Yaffe, {\it Confinement in {SU(N)} lattice gauge theories},  {\em Phys.
  Rev.} {\bf D21} (1980) 1574.

\bibitem{Mack:1979rq}
G.~Mack and V.~B. Petkova, {\it Comparison of lattice gauge theories with gauge
  groups {{Z(2)}} and {SU(2)}},  {\em Ann. Phys.} {\bf 123} (1979) 442.

\bibitem{Mack:1980kr}
G.~Mack and V.~B. Petkova, {\it Sufficient condition for confinement of static
  quarks by a vortex condensation mechanism},  {\em Ann. Phys.} {\bf 125}
  (1980) 117.

\bibitem{Tomboulis:1981vt}
E.~Tomboulis, {\it The '{t} {H}ooft loop in {SU(2)} lattice gauge theories},
  {\em Phys. Rev.} {\bf D23} (1981) 2371.

\bibitem{Tomboulis:1993wm}
E.~T. Tomboulis, {\it Confinement via dynamical monopoles},  {\em Phys. Lett.}
  {\bf B303} (1993) 103--108.

\bibitem{Cea:2000rj}
P.~Cea and L.~Cosmai, {\it Magnetic condensation and confinement in lattice
  gauge theory},  {\em Nucl. Phys. Proc. Suppl.} {\bf 94} (2001) 486--489,
  [\href{http://xxx.lanl.gov/abs/hep-lat/0010034}{{\tt hep-lat/0010034}}].

\bibitem{Kadanoff:1971kz}
L.~P. Kadanoff and H.~Ceva, {\it Determination of an opeator algebra for the
  two-dimensional {Ising} model},  {\em Phys. Rev.} {\bf B3} (1971) 3918--3938.

\bibitem{Fradkin:1978th}
E.~H. Fradkin and L.~Susskind, {\it Order and disorder in gauge systems and
  magnets},  {\em Phys. Rev.} {\bf D17} (1978) 2637.

\bibitem{Davis:2000kv}
A.~C. Davis, T.~W.~B. Kibble, A.~Rajantie, and H.~Shanahan, {\it Topological
  defects in lattice gauge theories},  {\em JHEP} {\bf 11} (2000) 010,
  [\href{http://xxx.lanl.gov/abs/hep-lat/0009037}{{\tt hep-lat/0009037}}].

\bibitem{Cea:1997ff}
P.~Cea, L.~Cosmai, and A.~D. Polosa, {\it The lattice {Schr{\"o}dinger}
  functional and the background field effective action},  {\em Phys. Lett.}
  {\bf B392} (1997) 177--181,
  [\href{http://xxx.lanl.gov/abs/hep-lat/9601010}{{\tt hep-lat/9601010}}].

\bibitem{Cea:1997ku}
P.~Cea and L.~Cosmai, {\it Lattice background effective action: A proposal},
  {\em Nucl. Phys. Proc. Suppl.} {\bf 53} (1997) 574--577,
  [\href{http://xxx.lanl.gov/abs/hep-lat/9607015}{{\tt hep-lat/9607015}}].

\bibitem{Cea:1999gn}
P.~Cea and L.~Cosmai, {\it Probing the non-perturbative dynamics of {SU(2)}
  vacuum},  {\em Phys. Rev.} {\bf D60} (1999) 094506,
  [\href{http://xxx.lanl.gov/abs/hep-lat/9903005}{{\tt hep-lat/9903005}}].

\bibitem{Luscher:1992an}
M.~L{\"u}scher, R.~Narayanan, P.~Weisz, and U.~Wolff, {\it The
  {Schr{\"o}dinger} functional: A renormalizable probe for non {Abelian} gauge
  theories},  {\em Nucl. Phys.} {\bf B384} (1992) 168--228,
  [\href{http://xxx.lanl.gov/abs/hep-lat/9207009}{{\tt hep-lat/9207009}}].

\bibitem{Luscher:1995vs}
M.~L{\"u}scher and P.~Weisz, {\it Background field technique and
  renormalization in lattice gauge theory},  {\em Nucl. Phys.} {\bf B452}
  (1995) 213--233, [\href{http://xxx.lanl.gov/abs/hep-lat/9504006}{{\tt
  hep-lat/9504006}}].

\bibitem{Sint:1994un}
S.~Sint, {\it On the {Schr{\"o}dinger} functional in {QCD}},  {\em Nucl. Phys.}
  {\bf B421} (1994) 135--158,
  [\href{http://xxx.lanl.gov/abs/hep-lat/9312079}{{\tt hep-lat/9312079}}].

\bibitem{Luscher:1993zx}
M.~Luscher, R.~Sommer, U.~Wolff, and P.~Weisz, {\it Computation of the running
  coupling in the {SU(2)} {Yang-Mills} theory},  {\em Nucl. Phys.} {\bf B389}
  (1993) 247--264, [\href{http://xxx.lanl.gov/abs/hep-lat/9207010}{{\tt
  hep-lat/9207010}}].

\bibitem{Rossi:1980jf}
G.~C. Rossi and M.~Testa, {\it The structure of {Yang-Mills} theories in the
  temporal gauge. 1. general formulation},  {\em Nucl. Phys.} {\bf B163} (1980)
  109.

\bibitem{Rossi:1980pg}
G.~C. Rossi and M.~Testa, {\it The structure of {Yang-Mills} theories in the
  temporal gauge. 2. perturbation theory},  {\em Nucl. Phys.} {\bf B176} (1980)
  477.

\bibitem{Gross:1981br}
D.~J. Gross, R.~D. Pisarski, and L.~G. Yaffe, {\it {QCD} and instantons at
  finite temperature},  {\em Rev. Mod. Phys.} {\bf 53} (1981) 43.

\bibitem{Efron:1982}
B.~Efron, {\em Jackknife, the Bootstrap and Other Resampling Plans}.
\newblock SIAM Press, Philadelphia, 1982.

\bibitem{Shao:1995}
J.~Shao and D.~Tu, {\em The Jackknife and the Bootstrap}.
\newblock Springer, New York, 1995.

\bibitem{Cea:2001nm}
P.~Cea and L.~Cosmai, {\it Abelian and center vortex condensation in {SU(3)}
  lattice gauge theory},  \href{http://xxx.lanl.gov/abs/hep-lat/0110002}{{\tt
  hep-lat/0110002}}.

\end{thebibliography}
\providecommand{\href}[2]{#2}\begingroup\raggedright\endgroup

\end{document}